\documentclass[a4paper,fleqn,usenatbib]{mnras}

\usepackage[T1]{fontenc}
\usepackage{ae,aecompl}
\pdfoutput=1

\usepackage{graphicx}	
\usepackage{amsmath}	
\usepackage{amssymb}	

\usepackage{color}
\usepackage{float}
\usepackage{txfonts}
\usepackage{units}
\usepackage{url}

\newcommand{\suzaku}{\textit{Suzaku}~}
\newcommand{\lsi}{LS~I~+61$^{\circ}$303}
\newcommand{\grs}{GRS~1915+105}
\newcommand{\vcyg}{V404~Cyg}

\title[X-ray and radio QPOs in \lsi{}]{Hour time-scale QPOs in the X-ray and radio emission of \lsi{}}

\author[S. N\"osel et al.]{S.~N\"osel,$^{1}$
 R.~Sharma,$^{1}$\thanks{E-mail: rsharma@mpifr-bonn.mpg.de}
 M.~Massi,$^{1}$ 
 G.~Cim\`o$^{2,3}$ and
 M.~Chernyakova$^{4,5}$
\\
$^{1}$Max-Planck-Institut f\"ur Radioastronomie, Auf dem H\"ugel 69, D-53121 Bonn, Germany \\
$^{2}$Joint Institute for VLBI ERIC, P.O. Box 2, 7990 AA Dwingeloo, The Netherlands \\
$^{3}$Netherlands Institute for Radio Astronomy, P.O. Box 2, 7990 AA Dwingeloo, The Netherlands \\
$^{4}$School of Physical Sciences and C-fAR, Dublin City University, Dublin 9, Ireland \\
$^{5}$Dublin Institute for Advanced Studies, 31 Fitzwilliam Place, Dublin 2, Ireland
}

\date{Accepted XXX. Received YYY; in original form ZZZ}

\pubyear{2017}

\begin{document}
\label{firstpage}
\pagerange{\pageref{firstpage}--\pageref{lastpage}}
\maketitle

\begin{abstract}

\lsi{} is an X-ray binary with a radio outburst every $\sim$27 days. Previous studies of the stellar system revealed radio microflares superimposed on the large radio outburst.
We present here new radio observations of \lsi{} at 2.2~GHz with the Westerbork Synthesis Radio Telescope (WSRT). Using various timing analysis methods we find significant Quasi-Periodic Oscillations (QPOs) of 55~minutes stable over the duration of 4~days.
We also use archival data obtained from the \suzaku satellite 
at X-ray wavelengths. We report here for the first time significant X-ray QPOs of about 2~hours present over the time span of 21~hours. We compare our results with the previously reported QPO observations 
and we conclude that the QPOs seem to be associated with the radio outburst, 
 independent of the amplitude of the outburst.
Finally, the different QPO time-scales are discussed in the context of magnetic reconnection.



\end{abstract}

\begin{keywords}
Radio continuum: stars -- X-rays: binaries -- X-rays: individual (\lsi{})\end{keywords}




\section{Introduction}

Microquasars are excellent laboratories to probe the accretion-ejection coupling in accreting compact objects. Microquasars are radio emitting X-ray binaries, i.e., stellar systems where a compact object (a neutron star or a black hole) accretes matter from a normal star.
From the accretion disc and inflow, which emits in X-rays, a relativistic jet is launched which emits in the radio band \citep{Mirabel1999}.



Timing analysis of variability is a powerful tool to investigate variations around the accretion flow and their relationship with variations in the jet.
In fact, in one microquasar, the black hole X-ray binary \grs, a link between Quasi-Periodic Oscillations (QPOs) in the accretion flow and in the jet has been observed. Radio oscillations on time-scales of 20--40  minutes have been clearly related to quasi-periodic dips in the X-ray light curve (see Fig. 5 in \citealt{Pooley1997}, Fig. 3 in \citealt{Mirabel1998}, Fig. 3 in \citealt{Klein-Wolt2002}).
\citet{Belloni1997} suggested that the X-ray dips correspond to the disappearance of the inner part of the accretion disc. The relation of the radio oscillations to these dips led to the hypothesis that at least some of the material from the inner disc 
had in fact been ejected from the system, instead of simply falling into the black hole (see references in \citealt{Mirabel1999}).

Variability with time-scales of minutes--hours has been observed only in few microquasars:
\vcyg{} \citep{Han1992}, Cygnus~X-1  \citep{Marti2001}, \grs{} (e.g. \citealt{ Rodriguez1997, fender97, Fender2002}) and \lsi{} (e.g. \citealt{Peracaula1997, Taylor1992}).
Among them, the binary stellar system \lsi{} with its 
strong and periodic radio outbursts
\citep{Gregory2002, MassiTorricelli2016} represents a favourable target for QPO observations.  
Previously, one detection of radio QPO \citep{Peracaula1997} and hints for X-ray QPO \citep{Harrison2000}  with time-scales of minutes--hours were observed in \lsi.

The aim of this work is to verify the suggested presence of QPOs with time-scales of minutes--hours in the X-ray emission of \lsi{} and corroborate the presence of radio QPOs with the same time-scales. 

\begin{figure}
\begin{center}
\includegraphics[width=1\columnwidth]{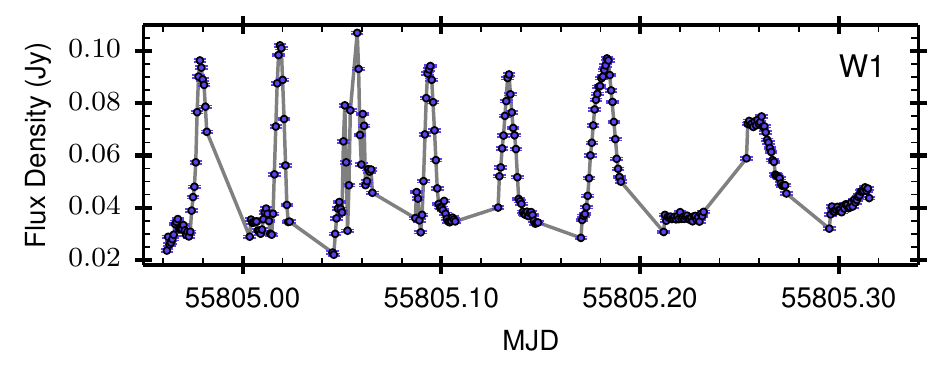}
\includegraphics[width=1\columnwidth]{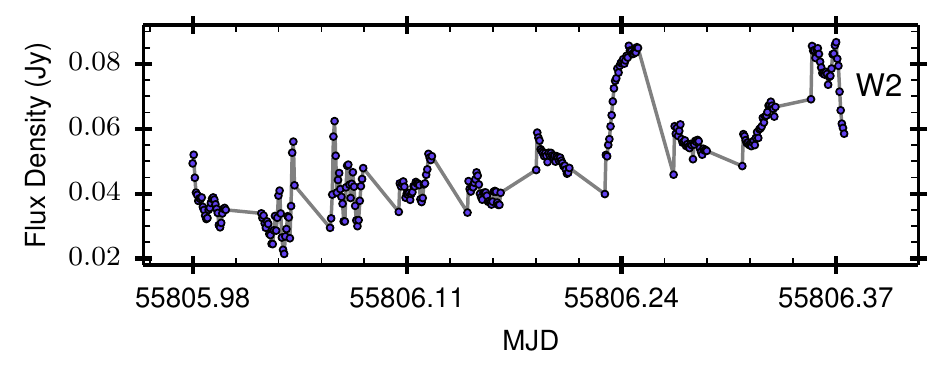}
\includegraphics[width=1\columnwidth]{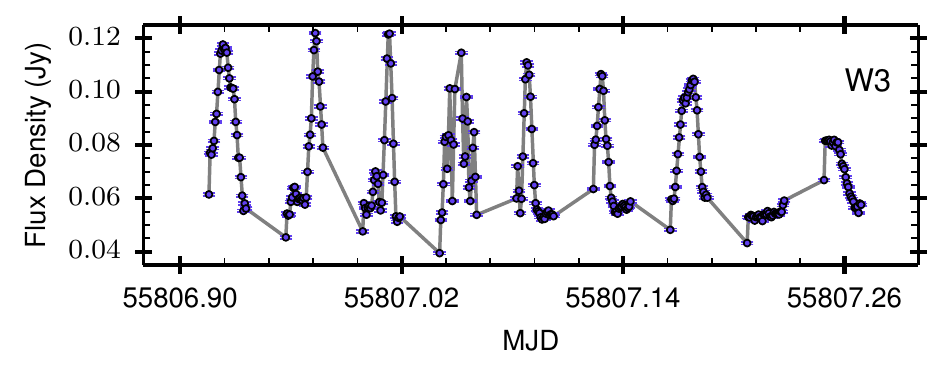}
\includegraphics[width=1\columnwidth]{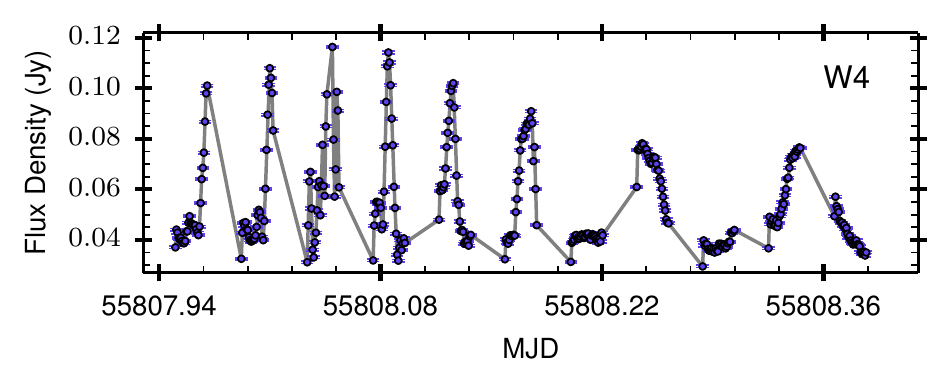}
\caption{Top to bottom: Radio light curves of \lsi{} at 2.2 GHz for all four days of observations with WSRT. Distinct microflares are already visible at all days with 7, 8 and 8 microflares for observations W1, W3 and W4, respectively. Observation W2 show hints of small microflares. The grey line is used to connect the data points.}
\label{fig:lightcurves_WSRT_Sband}
\end{center}
\end{figure}

\begin{figure}
\begin{center}
\includegraphics[width=1\columnwidth]{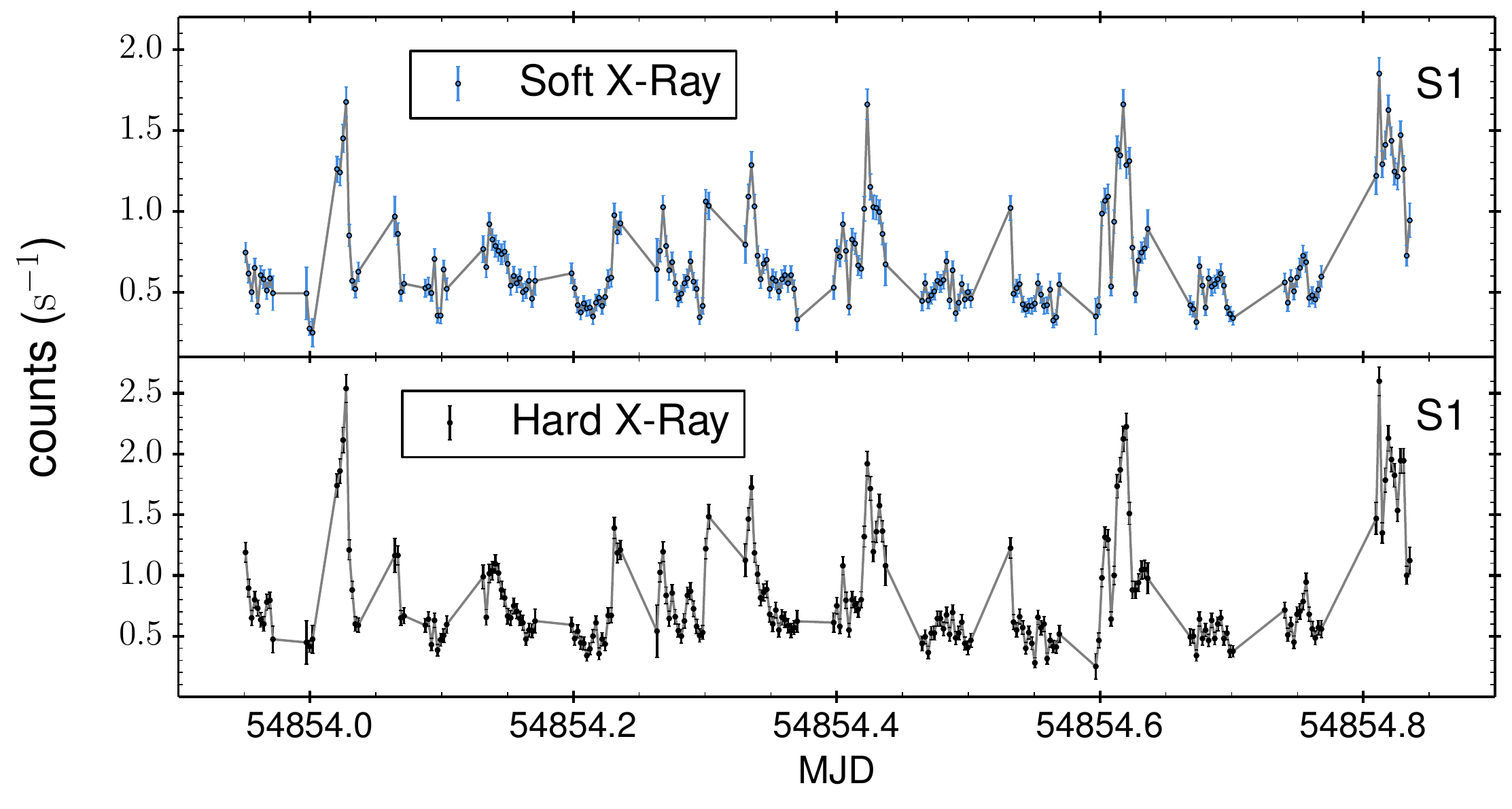} 
\includegraphics[width=1\columnwidth]{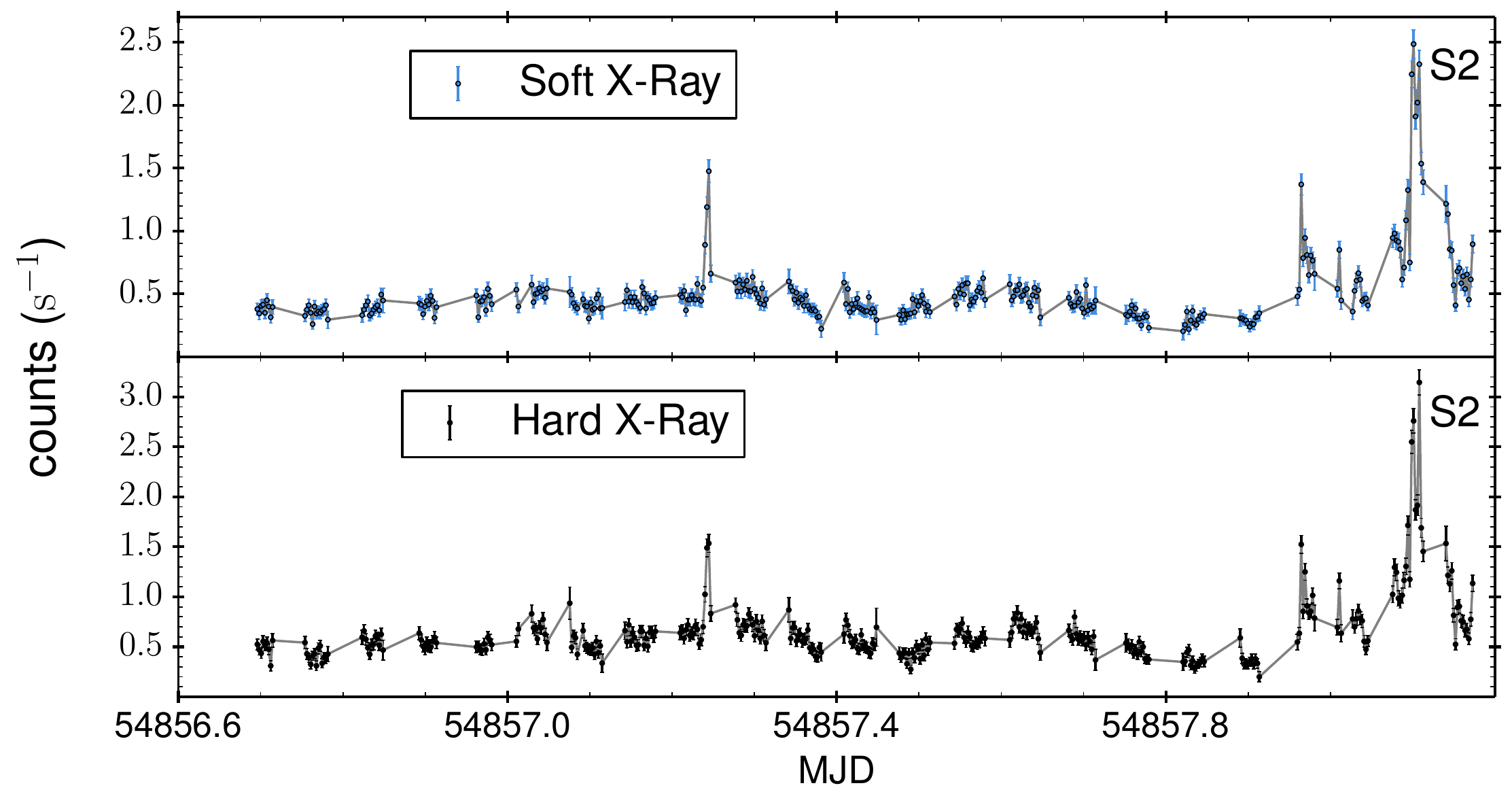}
\includegraphics[width=1\columnwidth]{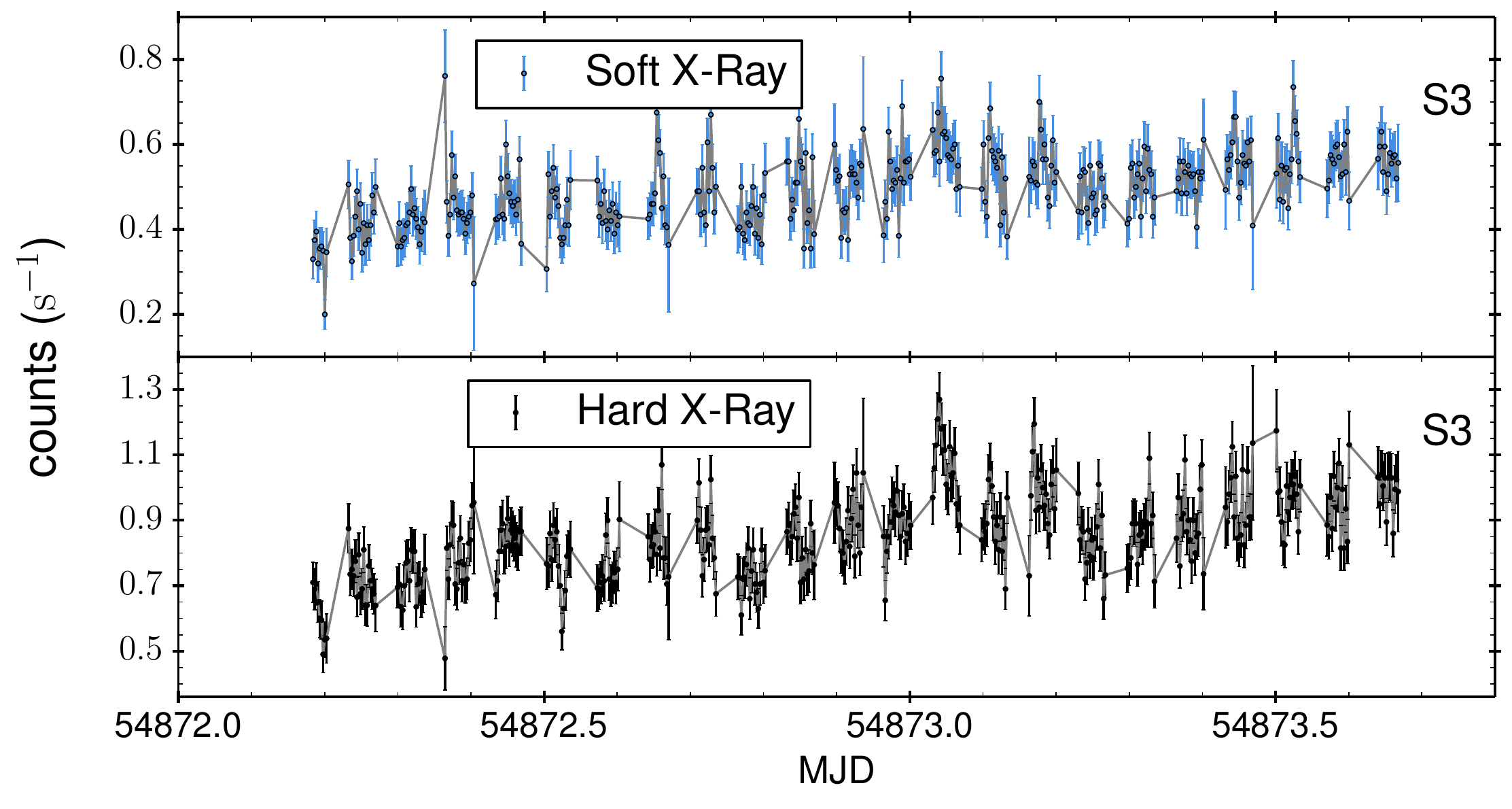} 
\caption{Top to bottom: X-ray light curves of \lsi{} for observations done with the \suzaku satellite. Soft X-ray data are represented in blue colour and hard X-ray data in black colour. The grey line is used to connect the data points. Timing analysis reveals QPOs only for observation S1.}
\label{fig:lightcurves_suzaku_all}
\end{center}
\end{figure}

\subsection{Previously reported variability in \lsi}

The stellar binary system \lsi{} is a high-mass X-ray binary (HMXB) comprising of a massive star that shows optical spectra of a B0 V star \citep{Casares2005} and a compact object.
As shown in \citet{Massi2017}, \lsi{} fits the trend of  intermediate luminosity accreting black holes, the systems
where the X-ray spectrum is thought to originate from inverse Compton processes in their inner accretion flow \citep{Yang2015b,Yang2015a}.
\lsi{} is located at a distance of $2.0 \pm 0.2$ kpc \citep{Frail1991} and has an orbital period, $P_{\text{orb}}=$  $26.4960 \pm 0.0028$ days \citep{Gregory2002}. 
Its orbital phase is defined as $\Phi_{\text{orb}}= \dfrac{t-t_0}{P_{\text{orb}}} - \textrm{int} \left(\dfrac{t-t_0}{P_{\text{orb}}} \right)$, where,  
$t_0=$  MJD 43366.275.
The source has a radio outburst every $26.70 \pm 0.05$ days \citep{Ray1997, Jaron2013}. 
The radio outburst in \lsi{} is also modulated over a long-term period, $P_{\text{long}}= 1626 \pm 68 $ days \citep{MassiTorricelli2016}. 
The superorbital phase is defined analogous to the orbital phase as, $\Theta= \dfrac{t-t_0}{P_{\text{long}}} - \textrm{int} \left(\dfrac{t-t_0}{P_{\text{long}}} \right)$.

Previous studies of this source have revealed short-term radio variability superimposed on the radio outburst. For the first time, a step-like pattern was observed in \lsi{} with the Westerbork Synthesis Radio Telescope (WSRT) phased array with variability time-scale of $\sim10^3$ seconds during the decay of the radio outburst \citep{Taylor1992}. The first systematic analysis of short-term variability in \lsi{} was performed using VLA observations by \citet{Peracaula1997}, wherein microflares with recurrence period of 1.4 hours were reported to be active for $\sim$8 hours with an amplitude of $\sim$4~mJy during the decay of the observed outburst. 
At higher energy, \citet{Harrison2000} indicated $\sim$30--40 minutes variability in one of the X-ray observations done with the Advanced Satellite for Cosmology and Astrophysics (ASCA) and associated with the onset of a radio outburst. 
Longer time-scale radio QPOs of 15.4~hours and 2~days were observed by 
\citet{jaron2017} and \citet{Zimmermann2015}, respectively.

In Sect.~2, we describe new observations and their data analysis. The results of the analysis are presented in Sect.~3. In Sect.~4, we compare our results with the previous QPO observations of the source. Finally, we conclude and discuss our results in Sect.~5.

\setcounter{table}{0}
\begin{table*}
\caption{Log of WSRT and \suzaku data used in this work.}
\centering
\begin{tabular}{lcccccc}
\hline
Telescope  & Obs. starting date &  Label &  MJD &  $\Phi_{\text{orb}}$ & $\Theta$ & $P_{\textrm{QPO}}$  \\ \hline 
WSRT (2.2 GHz) & 31.08.2011--03.09.2011 & W1--W4$^*$ & 55804.941--55807.929 & 0.45--0.57  & 0.650--0.652 & 55.3 $\pm$ 0.3 min \\
\suzaku & 22.01.2009 & S1 & 54853.951 & 0.56& 0.065 & 2.4 $\pm$ 0.3 h \\
 (0.3--12 keV and & 25.01.2009 & S2 & 54856.696 & 0.67 & 0.067 & No QPOs \\
13--600 keV) & 10.02.2009 & S3 & 54872.184 & 0.25 & 0.076 & No QPOs \\ \hline 
\multicolumn{4}{c}{*Note: The MJD values for W1--W4 are 55804.941, 55805.959, 55806.895 and 55807.929.} 
\end{tabular}
\label{table:log}
\end{table*}

\setcounter{table}{1}
\begin{table*}
\caption{An overview of the previously reported QPOs in \lsi. The references are: $^\textrm{a}$\citet{jaron2017},  $^\textrm{b}$\citet{Peracaula1997}, $^\textrm{c}$\citet{Taylor1992} and $^\textrm{d}$\citet{Harrison2000}. The occurrence of the QPOs with respect to the radio outburst evolution is
given in the last column.}
\centering
\begin{tabular}{lccccccc}
\hline
Telescope  & Obs. starting date &  Label &  MJD &  $\Phi_{\text{orb}}$ & $\Theta$ & $P_{\textrm{QPO}}$ & Comments \\ \hline 
Effelsberg $^\textrm{a}$  & 17.04.2014 & J1 & 56764.726 & 0.68 & 0.240 &  15.4 $\pm$ 0.6 h & Decay of radio outburst.\\
(4.85, 8.35, 10.45 GHz) & & & & & \\
VLA$^\textrm{b}$ (5 GHz)& 06.06.1990 & P1 & 48048.292 & 0.72 & 0.880 & 1.4 h ? & Less significant QPOs during decay.\\
  &  09.09.1993 & P2 & 49239.042 & 0.66 & 0.612 &1.4 h & Decay of radio outburst.\\
  & 13.09.1993 & P3 & 49243.042 & 0.81 & 0.614 & No QPOs & High flux quiescence. \\
EVN $^\textrm{c}$ 
(5 GHz) & 01.10.1987 & T1 & 47069.875 & 0.78 & 0.278 & 1000 s & Decay of radio outburst. 
\\
ASCA$^\textrm{d}$ (1--5 keV) & 03.02.1994 & H1 & 49386.104 & 0.20 & 0.702 & No QPOs & Before onset of radio outburst.\\
 & 09.02.1994 & H2 & 49392.312 & 0.43 & 0.706 & 30--40 min & Onset of radio outburst.\\ \hline
\end{tabular}
\label{table:overview}
\end{table*}

\section{Observations and Data analysis}

\subsection{New WSRT radio observations}

The here presented WSRT observations of \lsi{} at 2.2~GHz are part of a larger
observation whose data reduction is still in progress. 
The source was observed for four consecutive days (see Table~\ref{table:log}) from August 31, 2011 until September 3, 2011.
We obtained 9--11 scans each of 30 minute length.
The integration time was 1~minute.
The data reduction of the raw data from the telescope was performed using
the Common Astronomy Software Package\footnote{For detailed information about CASA visit \url{https://casa.nrao.edu}.}~(CASA). We used 3C48 as a flux calibrator. The obtained light curves are shown in Fig.~\ref{fig:lightcurves_WSRT_Sband} for all four days.

\subsection{X-ray observations with \suzaku}

For X-ray observations of \lsi{} we use the archival data obtained from the \suzaku satellite \citep{Chernyakova2017}. The data consist of three  observation runs in the year 2009 (see Table~\ref{table:log}), where each observation run includes data of soft X-ray (0.3--12~keV) and hard X-ray (13--600~keV). The total duration of observation for each run varies, with 21.3~hours for S1, 35.5~hours for S2 and 35.6~hours for S3. 
The data reduction was done using  the \texttt{HEASOFT} \footnote{\url{http://heasarc.gsfc.nasa.gov/docs/software/lheasoft/}.}
v.6.16 software package and the spectral modelling was performed in the \texttt{XSPEC} environment
v.12.8.2. For more details on the data reduction see Sect.~2.1
in \citet{Chernyakova2017}.
The data were integrated over 200~seconds.
The  obtained light curves are shown in Fig.~\ref{fig:lightcurves_suzaku_all}.


\subsection{Timing analysis}
To find variability on short time-scales in \lsi{} at both radio and X-ray wavelengths, we use three timing analysis techniques.
The data were initially analysed using the Lomb-Scargle periodogram which is a powerful tool to find and test the significance of weak periodic signals in unevenly sampled data \citep{Lomb1976, Scargle1982}.
The significance of the found periodic signal is tested with the Fischer-randomization test where the flux is permuted thousand times and thousand new
randomised time series are created and their periodograms calculated \citep{Nemec1985}.
The proportion of permuted time series that contain a higher peak in the periodogram than the original
periodogram at any frequency then gives the false alarm probability, $p$ of the peak. If $p < 0.01$,
the period is significant and if $0.01 < p < 0.1$ the period is marginally significant.
We then analysed the data using the phase dispersion minimization (PDM) method which is useful when we have a limited number of data points available and the
light curve is non-sinusoidal \citep{Stellingwerf1978}.
In this technique, the data-series is folded on many trial
frequencies. The folded data are then divided into various bins and their variance is calculated. The PDM
statistics is the ratio of overall variance of different bins and the variance of the original data. If the trial
period is a true period, then we observe a local minimum in the PDM which approaches zero. The minima
in the PDM must be an approximate mirror image of the maxima in the Lomb-Scargle periodogram.
Finally, the data were folded in phase averaged bins with the significant period, $P$, found from both the above explained methods. The phase $\phi$ is calculated as 
$
\phi = \dfrac{t-t_0}{P} - \textrm{int} \left(\dfrac{t-t_0}{P} \right)
$, 
where, $t_0$ is the first data point of each observation.

\section{Results}

\subsection{Variability in radio}

\begin{figure}
\begin{center}
\includegraphics[width=1\columnwidth]{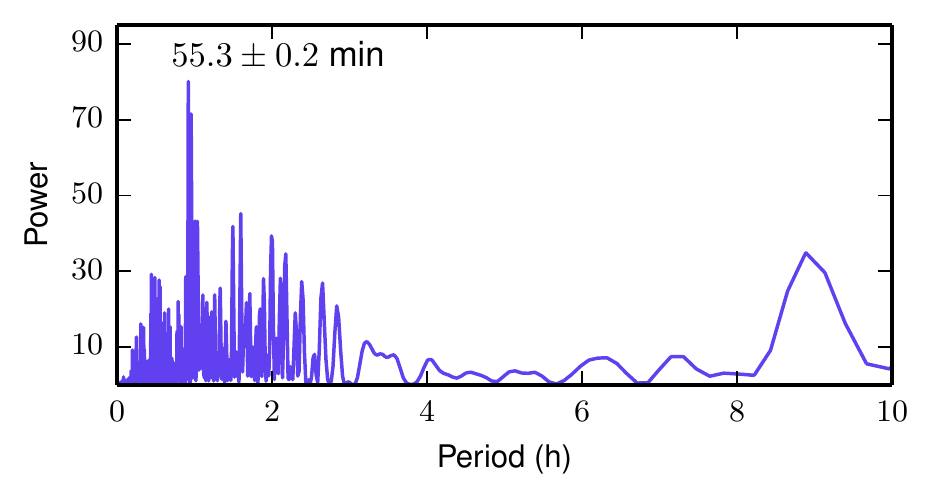} 
\includegraphics[width=1\columnwidth]{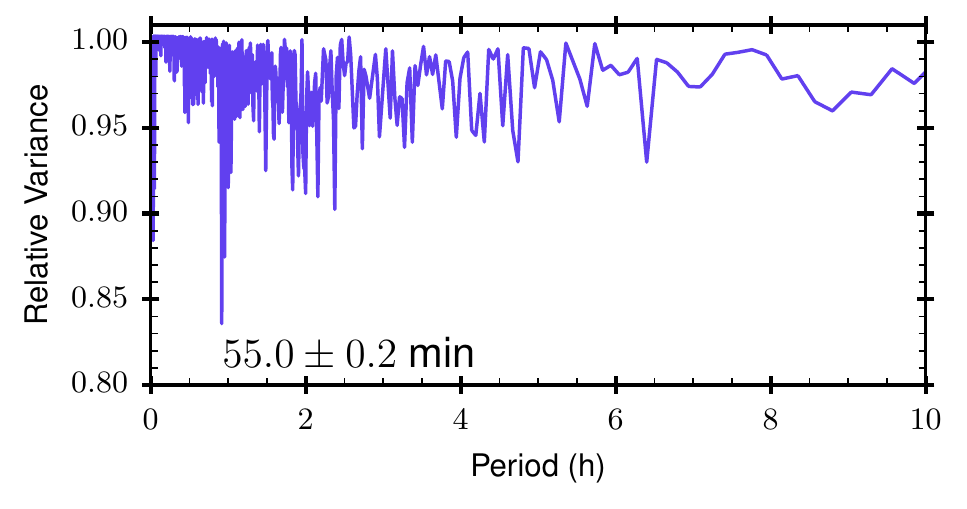} 
\includegraphics[width=1\columnwidth]{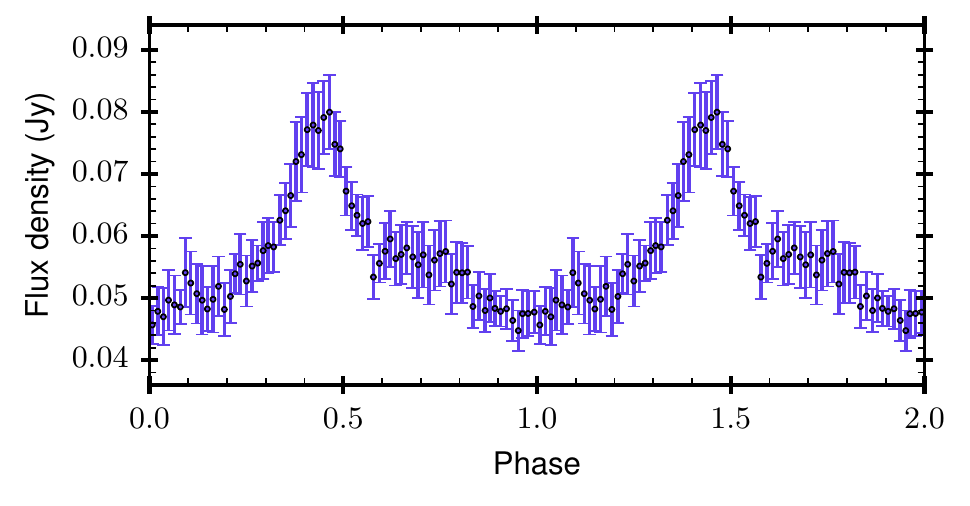}
\caption{Timing analysis results of the radio (2.2 GHz) data observed with the WSRT for all four days. Top: Lomb-Scargle periodogram. Middle: The true period yields the minimum relative variance in the PDM curve. Bottom: Phase averaged data folded with the significant period of 55.3 minutes and with a phase bin size of 0.014. For clarity, the data are repeated in the second cycle.}
\label{fig:timing_analysis_WSRT}
\end{center}
\end{figure}

The light curves for the 2.2~GHz data (see Fig.~\ref{fig:lightcurves_WSRT_Sband}) show well sampled microflares with varying amplitude and width. Observations W1, W3 and W4 show 7, 8 and 8 distinct microflares, respectively. 
Observation W2 shows only hints of small microflares.
The Lomb-Scargle periodogram of the combined dataset of four days 
is shown in the top panel of Fig.~\ref{fig:timing_analysis_WSRT}. We find a significant periodic signal in the data with a period of $55.3 \pm 0.2$~minutes. The PDM statistics (see middle panel of Fig.~\ref{fig:timing_analysis_WSRT}) gives a significant period of $55.0 \pm 0.2$~minutes. Finally, the phase averaged data folded with the found period from the periodogram are shown in the bottom panel of Fig.~\ref{fig:timing_analysis_WSRT}. 
The well clustered data peaking at one phase confirms the presence of the period of 55.3 minutes over four days.

\begin{figure}
\begin{center}
\includegraphics[width=1\columnwidth]{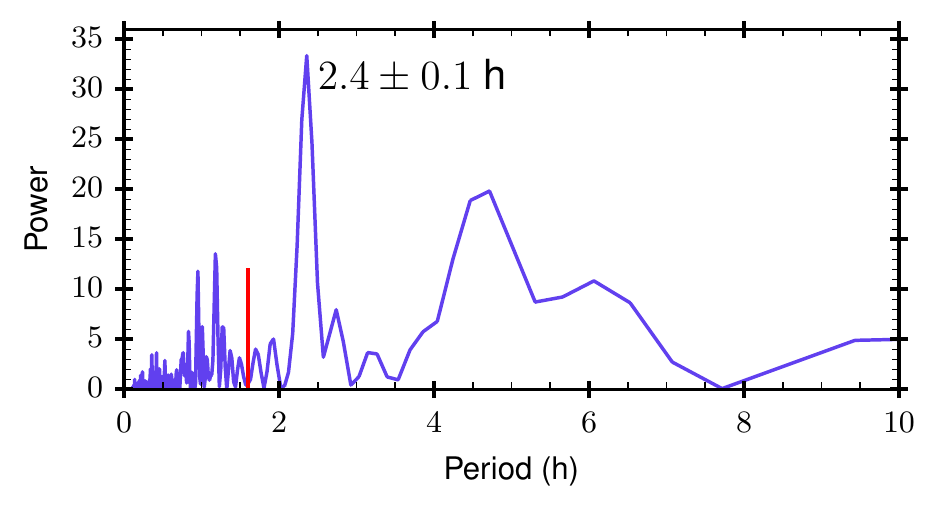} 
\includegraphics[width=1\columnwidth]{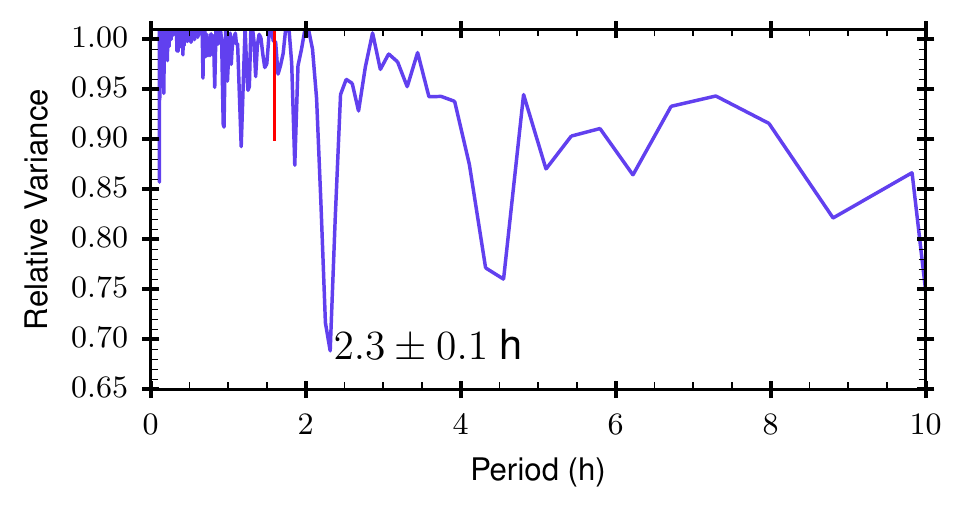} 
\includegraphics[width=1\columnwidth]{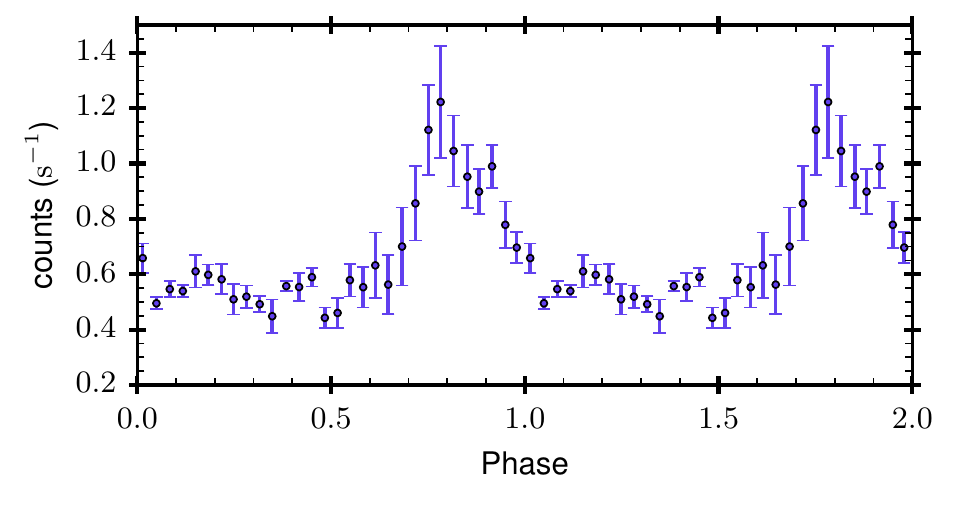}
\caption{Timing analysis results of the soft X-ray (0.3--12 keV) data observed with the \suzaku satellite for the first run of the observation session~(S1). Top: Lomb-Scargle periodogram. The red line indicates the orbital period  of 1.6 hours of the satellite. Middle: The true period yields the minimum relative variance in PDM and is specular of the peak in Lomb-Scargle periodogram. Bottom: Phase averaged data folded with the significant period of 2.4 hours and with a phase bin size of 0.033. For clarity, the data are repeated in the second cycle.}
\label{fig:timing_analysis_suzaku}
\end{center}
\end{figure}

\subsection{Variability in X-rays}

The light curves for both soft and hard X-ray (see Fig.~\ref{fig:lightcurves_suzaku_all}) show at least 7 distinct flares for S1. The observation S2 was done three days later and shows one isolated flare followed by the other two flares.
The observation S3 which was done 19 days later has hardly any visible flares.
The timing analysis results of soft and hard X-ray data are similar and therefore we show the results of soft X-rays only.
For S1, the Lomb-Scargle periodogram is shown in the top panel of Fig.~\ref{fig:timing_analysis_suzaku} with a significant period of $2.4 \pm 0.1$ hours. 
The PDM statistics (see the middle panel of Fig.~\ref{fig:timing_analysis_suzaku}) reveal a significant period of $2.3 \pm 0.1$ hours. 
The red line in the top and in the middle panel of Fig.~\ref{fig:timing_analysis_suzaku} indicates the orbital period of 1.6 hours of the \suzaku satellite, which does not interfere with our results.
Finally, the phase averaged data folded with the found period are shown in the bottom panel of Fig.~\ref{fig:timing_analysis_suzaku}. The data are well clustered at one phase which confirms the period of 2.4 hours. For S2 and S3, no QPOs were found during the analysis.

\begin{figure*}
\begin{center}
\includegraphics[width=1\columnwidth]{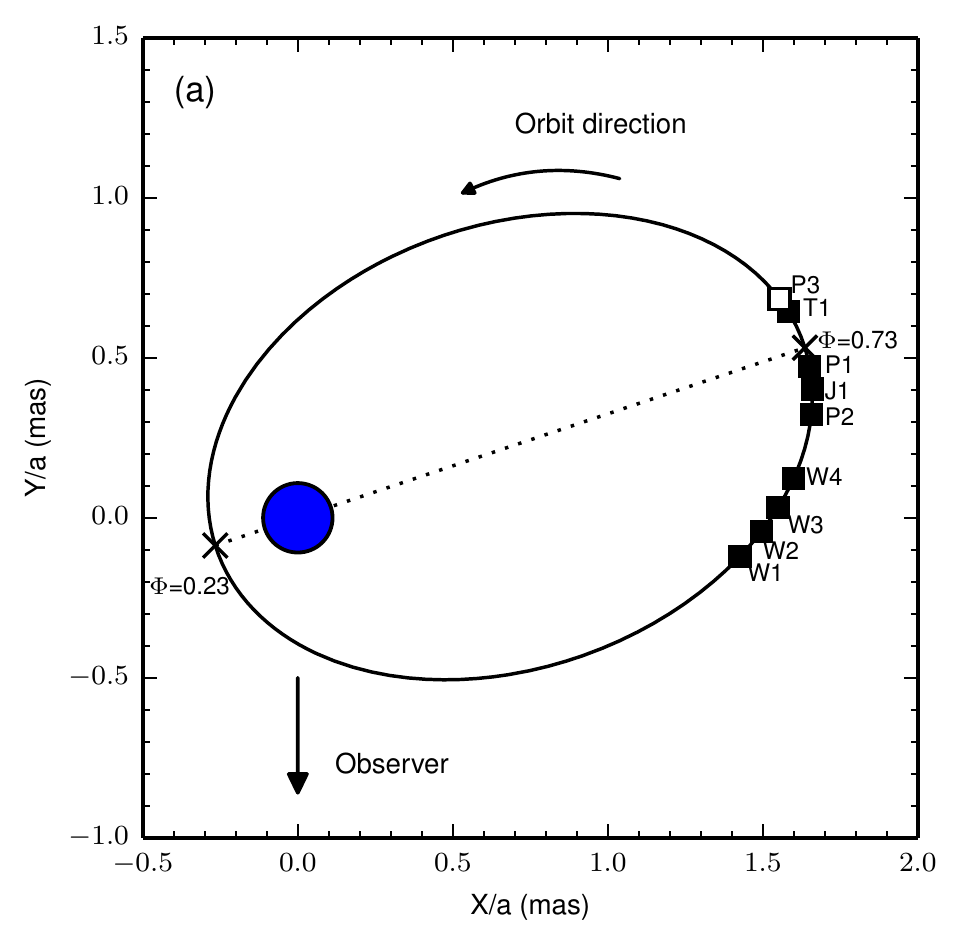}
\includegraphics[width=1\columnwidth]{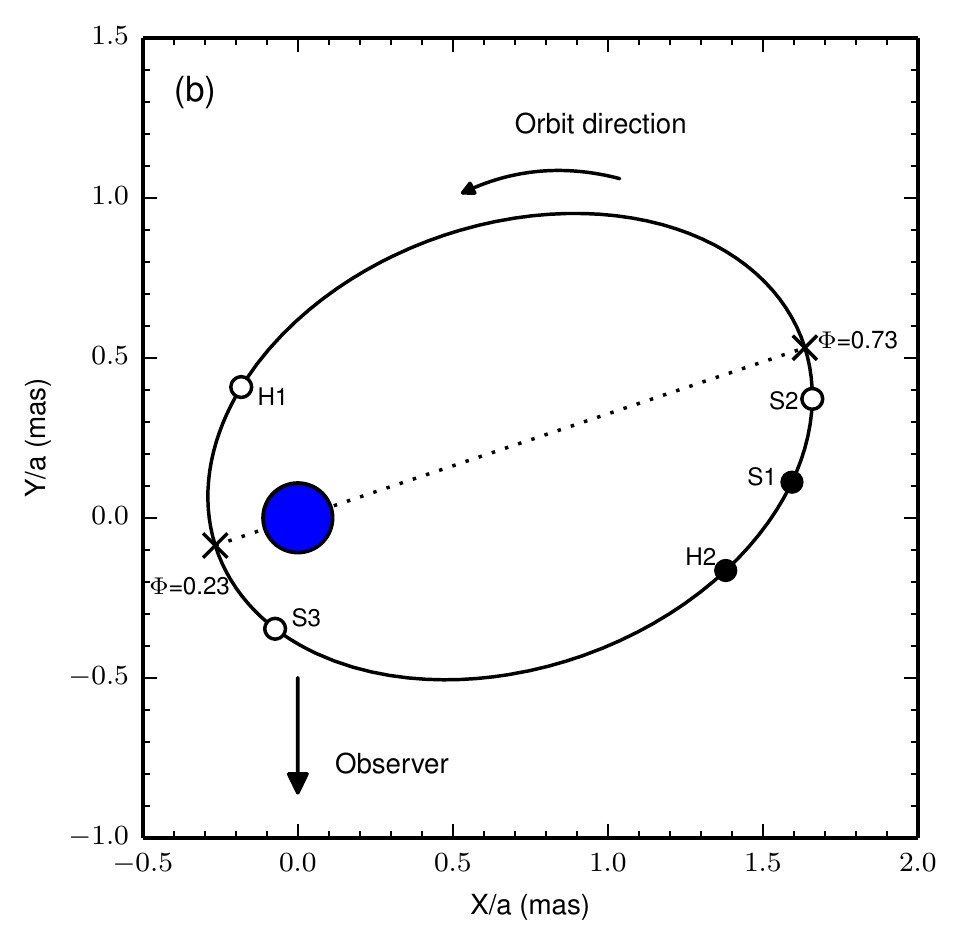} 
\caption{(a) Radio observations (squares) and (b) X-ray observations (circles) as given in Table~\ref{table:log} and Table~\ref{table:overview} marked here 
along the relative orbit of the compact object in \lsi{}. The crosses mark the periastron and apoastron phase intervals. The QPO detections are represented by black squares and circles, and non-detections are represented by open squares and circles. The orbital parameters are  taken from \citet{Casares2005} with the Be star at ellipse focus (0,0). }
\label{fig:QPO_orbit}
\end{center}
\end{figure*}

\begin{figure}
\begin{center}
\includegraphics[width=1\columnwidth]{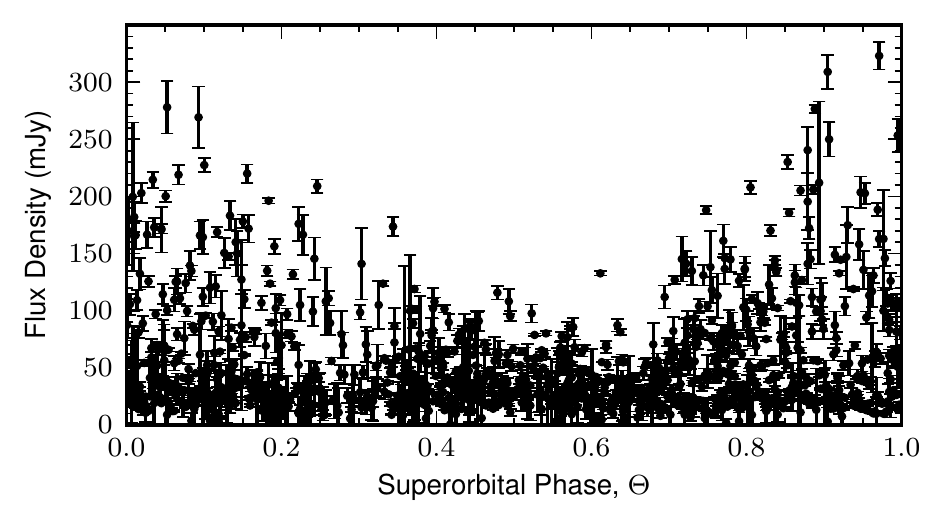} 
\includegraphics[width=1\columnwidth]{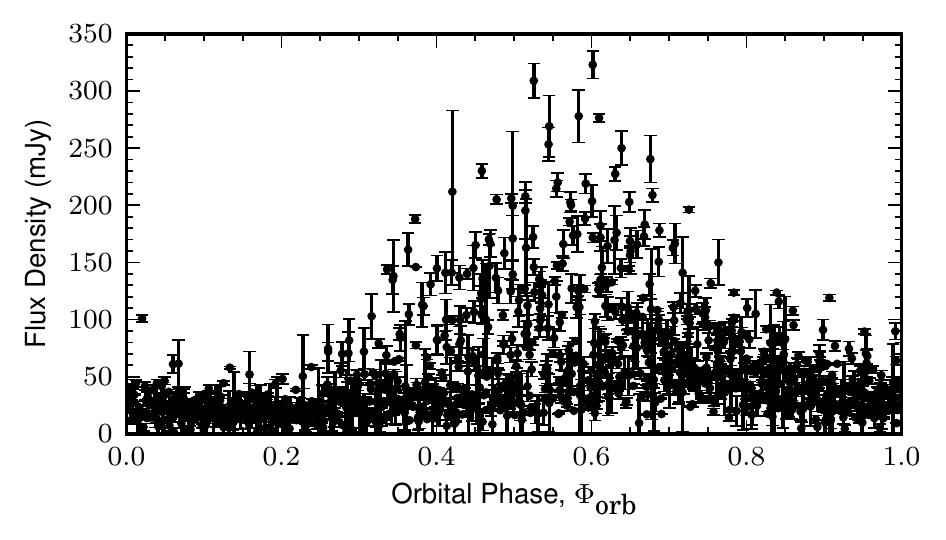} 
\caption{Radio data of \lsi{} of $\sim$37 years from \citet{MassiTorricelli2016}. Top: Radio data folded with the superorbital period of 1626 days. The radio flux modulates with minimum flux around $\Theta \sim$0.6 and maximum flux around $\Theta \sim$0.1. Bottom: Radio data folded  with the orbital period of 26.496 days with flux density peak observed at $\Phi_{\textrm{orb}} \sim$0.6.}
\label{fig:FluxVsPhase}
\end{center}
\end{figure}

\begin{figure}
\begin{center}
\includegraphics[width=1\columnwidth]{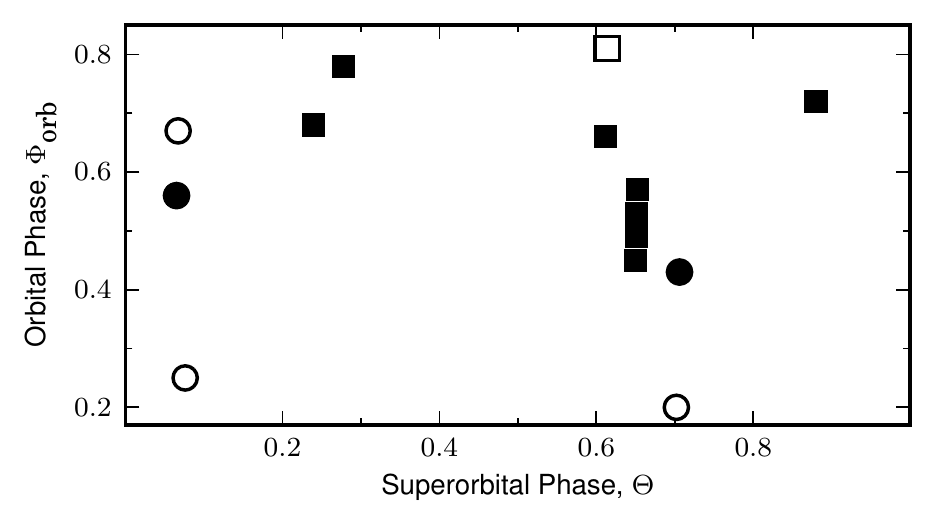} 
\caption{Radio and X-ray observations as in Table~\ref{table:log} and Table~\ref{table:overview} represented with their corresponding orbital phase and superorbital phase. The radio and X-ray observations are represented as squares and circles, respectively. The colour coding is same as in Fig.~\ref{fig:QPO_orbit}.}
\label{fig:QPO_ThetaVsPhi}
\end{center}
\end{figure}

\section{Comparison of our results with previous QPO observations}

In this work we report about two new observations of \lsi{}  with QPOs, one at radio and one at X-ray wavelengths.
In order to understand the implications of our results, we compare them with the previously reported QPOs. Table~\ref{table:overview} gives an overview of the observations done in the past with a comparison between their orbital phase, superorbital phase and period of QPO ($P_{\textrm{QPO}}$). In Fig.~\ref{fig:QPO_orbit}, we present the relative orbital geometry of \lsi{} as in \citet{Casares2005}. The observations of Table~\ref{table:log} and Table~\ref{table:overview} are shown along the orbit. Figure~\ref{fig:QPO_orbit}~(a) and Fig.~\ref{fig:QPO_orbit}~(b) show the radio and X-ray observations, respectively.
As shown in the lower panel of  Fig.~\ref{fig:FluxVsPhase} ($\sim$37 years of radio data, \citealt{MassiTorricelli2016}), the FWHM of radio outburst
occurs in the orbital range 0.4--0.8 with its peak value around $\sim$ 0.6.
Figure~\ref{fig:QPO_orbit}~(a) shows that this orbital range (0.4--0.8) is  well sampled.
In fact, observation 
W1 is at phase 0.45 and observation P3 is at phase 0.81.
The good sampling  between the orbital phases 0.4--0.8,
indicates that radio QPOs are always associated with radio outbursts, either during their onset
or during their decay. Especially vital is the radio observation P3 (performed four days after the outburst peak, i.e., at the end of the radio decay phase) where the flux density
remained quiescent at a flux level of 55 mJy for a long interval of 10 hours and no QPOs were detected for the same.
X-ray observations are more sparse. In Fig.~\ref{fig:QPO_orbit}~(b), two of the X-ray observations, H1 and S3, were taken close to the periastron. Both the observations are of non-detection of QPOs. These orbital phases are also rather displaced from those where the radio outburst occur. 
Observation S2 is at phase 0.67, where statistically the radio outburst is on its decay. It is interesting that
the X-ray observation S2 shows  variability but no periodicity (i.e., no QPOs).
The only X-ray QPO detections, H2 and S1,
are at the  orbital phases, 0.43 and 0.56, respectively, i.e.,  well in the range where statistically
the onset of the radio outburst occurs.

The amplitude of the radio outburst of \lsi{} is modulated with the superorbital period. Top panel of Fig.~\ref{fig:FluxVsPhase} shows $\sim$37 years of radio data folded with the long-term period ($P_{\text{long}}$). The amplitude can change from the minimum flux density at $\Theta_{\text{min}} \sim$0.6 to the maximum flux density at $\Theta_{\text{max}} \sim$0.1 by a factor of 6.
Do we expect a dependence of QPO occurrence on the amplitude of the outburst? 
Figure~\ref{fig:QPO_ThetaVsPhi} illustrates the dependence of superorbital phase and orbital phase for all the reported observations of Table~\ref{table:log} and Table~\ref{table:overview}. 
Detections seem to be independent of $\Theta_{\text{min}}$ or $\Theta_{\text{max}}$, i.e., they are independent from the amplitude of the radio outbursts. 
This result would be  consistent with the precessing jet scenario of \lsi{} \citep{Massi2014}. 
In this scenario, the superorbital modulation is not related to an intrinsic flux density  enhancement
but only depends on the Doppler boosting effects (i.e., apparent flux density enhancement).

\section{Conclusions and discussion}
We presented here radio data of \lsi{} observed with the WSRT for four consecutive days in 2011. We also used the archival X-ray data of \lsi{} observed with the \suzaku  satellite in 2009 for three different epochs. The following are our conclusions:

\begin{enumerate}

\item[1.] Our timing analysis establishes the presence of significant radio QPOs of $\sim$55 minutes stable for the entire observation session of four days. This is the first time that short-time-scale radio QPOs are observed for such a long duration of time, in fact the $\sim$1.4 hours microflares observed by \citet{Peracaula1997} were observed for $\sim$8 hours. After the hints of X-ray variability in \lsi{} by \citet{Harrison2000}, we confirm here for the first time X-ray QPOs of $\sim$2~hours period over a duration of 21.3 hours. 

\item[2.] From the comparison between our work and previous QPO observations of \lsi,  there seems to be no dependence of the presence of QPOs on superorbital phase. 
Even though the good sampling of the orbital interval 0.4--0.8  allows us to localise the presence of 
radio QPOs to both the onset and the decay phase of the radio outbursts. The more sparse X-ray observations seem to indicate the presence of X-ray QPOs only 
during the onset of the radio outburst.

\item[3.] We further note that QPOs in \lsi{} show different periods (from $\sim$1000~seconds to $\sim$15~hours) at different epochs.

\end{enumerate}

In \grs, QPOs have been interpreted as ejection of relativistic plasma clouds or plasmoids (\citealt{Mirabel1999}, see their Fig.~7). 
Recently, magnetic reconnection events are invoked to explain  the ejection of plasmoids in the jets of blazars, wherein current sheet is fragmented into a chain of plasmoids. 
These plasmoids can be of different size and can be ejected with different time-scales (minutes, hours to days) (e.g. \citealt{Petropoulou2016, Sironi2016}). Such magnetic reconnection events can thus give rise to QPOs of different periods in the radio jet and can significantly modulate the main outburst.
Moreover, magnetic reconnection can also occur in the highly magnetised X-ray emitting accretion flow producing plasmoids and giving rise to QPOs at X-ray wavelengths. These plasmoids further expand adiabatically after leaving the accretion flow and could be seen as optically thin emission in the radio band \citep{Yuan2009}.
In this second scenario, i.e., plasmoids generated within the accretion flow, magnetic reconnection could be the physical process creating the delayed radio QPOs with respect to the X-ray QPOs of same time-scales, as observed in \grs{} (e.g.  \citealt{Mirabel1999, Klein-Wolt2002}). Future observations of \lsi{} should establish and provide a better understanding if the X-ray QPOs are present only during the onset of the radio outburst. Does this imply that X-ray QPOs are related solely to the ejection phase of plasmoids from the accretion flow? Are there associated delayed radio QPOs? Are the radio QPOs associated to the decay of the outburst still related to X-ray QPOs or are they just the result of magnetic reconnection occurring directly in the jet, as discussed above for blazars? These issues should be addressed by future observations.




\section*{Acknowledgements}
We thank Frederic Jaron and Sergio A. Dzib  for carefully reading the manuscript.
The Westerbork Synthesis Radio Telescope is operated by the ASTRON
(Netherlands Institute for Radio Astronomy) with support from the Netherlands
Foundation for Scientific Research (NWO). This research has made use of data
obtained from the \suzaku satellite, a collaborative mission between the
space agencies of Japan (JAXA) and the USA (NASA). This research has made use
of data and/or software provided by the High Energy Astrophysics Science
Archive Research Center (HEASARC), which is a service of the Astrophysics
Science Division at NASA/GSFC and the High Energy Astrophysics Division of
the Smithsonian Astrophysical Observatory.




\bibliographystyle{mnras}
\bibliography{reference} 
\bsp	
\label{lastpage}
\end{document}